\documentclass[sigconf]{acmart}

\AtBeginDocument{%
  }

\copyrightyear{2025}
\acmYear{2025}
\acmDOI{10.1145/3706598.3714191}
\setcopyright{cc}
\setcctype{by}

\acmConference[CHI '25]{CHI Conference on Human Factors in Computing Systems}{April 26-May 1, 2025}{Yokohama, Japan}
\acmBooktitle{CHI Conference on Human Factors in Computing Systems (CHI '25), April 26-May 1, 2025, Yokohama, Japan}
\acmISBN{979-8-4007-1394-1/25/04}

\begin{document}

\title{How Visualization Designers Perceive and Use Inspiration}

\author{Ali Baigelenov}
\email{abaigele@purdue.edu}
\orcid{0009-0003-6491-1874}
\affiliation{%
  \institution{Purdue University}
  \city{West Lafayette}
  \state{Indiana}
  \country{USA}
}

\author{Prakash Shukla}
\email{shukla37@purdue.edu}
\orcid{0009-0002-7416-1758}
\affiliation{%
  \institution{Purdue University}
  \city{West Lafayette}
  \state{Indiana}
  \country{USA}}

\author{Paul Parsons}
\email{parsonsp@purdue.edu}
\orcid{0000-0002-4179-9686}
\affiliation{%
  \institution{Purdue University}
  \city{West Lafayette}
  \state{Indiana}
  \country{USA}}

\renewcommand{\shortauthors}{Baigelenov et al.}

\begin{abstract}
  
  Inspiration plays an important role in design, yet its specific impact on data visualization design practice remains underexplored. This study investigates how professional visualization designers perceive and use inspiration in their practice. Through semi-structured interviews, we examine their sources of inspiration, the value they place on them, and how they navigate the balance between inspiration and imitation. Our findings reveal that designers draw from a diverse array of sources, including existing visualizations, real-world phenomena, and personal experiences. Participants describe a mix of active and passive inspiration practices, often iterating on sources to create original designs. This research offers insights into the role of inspiration in visualization practice, the need to expand visualization design theory, and the implications for the development of visualization tools that support inspiration and for training future visualization designers.
\end{abstract}




\received{12 September 2024}
\received[revised]{10 December 2024}
\received[accepted]{16 January 2025}

\maketitle

\section{Introduction}
The question of where creative ideas originate has been investigated across a wide variety of disciplines, including art, business, science, and design. Design, which has been defined as ``the ability to imagine that-which-does-not-yet-exist, to make it appear in concrete form as a new, purposeful addition to the real world'' \cite{nelson_design_2012}, inevitably involves the generation of new ideas, materials, and artifacts. Historically, different accounts of what design is have shaped the ways in which ideation and creativity have been studied. These accounts have ranged from viewing the designer as a `creative genius', in which their process is essentially a `black box' that is inseparable from the designer, to viewing the designer as an information processor, in which their process is a `glass box' that is entirely rational, objective, and separable from the designer \cite{fallman_design-oriented_2003}. Most current scholarship takes a more balanced approach, viewing design as personal and situated but also amenable to investigation and description \cite{schon_reflective_1983,fallman_design-oriented_2003,nelson_design_2012}. From this perspective, we can appreciate the role of the individual designer, including their training, experiences, and preferences, in shaping how they generate ideas, while also aiming to study their process to generate knowledge about design for researchers and practitioners. 

Virtually all design outcomes proceed from some combination of previous designs, phenomena in the world, and the personal experiences and knowledge of the designer \cite{eckert_sources_2000}. Design rarely proceeds `from scratch' and almost always involves one or more sources of inspiration. Researchers have investigated inspiration practices of designers across a variety of design disciplines, including knitwear design \cite{eckert_sources_2000}, interaction design \cite{chan_best_2015}, industrial and graphic design \cite{miller_searching_2014}, to name a few. Understanding how designers find and use sources of inspiration within a disciplinary context is beneficial, as it can help support creativity and hinder fixation \cite{miller_searching_2014}, identify which kinds of inspiration produce the best design ideas \cite{chan_best_2015}, provide vocabulary for thinking about new designs and describing designs to others \cite{eckert_sources_2000}, and better understand how idea generation happens \cite{goncalves_inspiration_2016}. 

To influence real-world design practice, design researchers must recognize practice as a central unit of analysis rather than simply a downstream beneficiary of research outcomes \cite{kuutti_turn_2014}. The visualization community has embraced more practice-led and practitioner-focused work in recent years, investigating how visualization practitioners work in the world (e.g., \cite{parsons_understanding_2022,alspaugh_futzing_2019,mckenna_design_2014,bigelow_reflections_2014,hoffswell_techniques_2020,mendez_bottom-up_2017,bako_understanding_2022}). However, topics relating to design cognition, such as ideation and inspiration use, have not yet been studied in detail \cite{parsons_design_2023}. The few related examples include work by Bako and colleagues on how visualization designers find and use visualization examples  \cite{bako_understanding_2022} and their effects on design outcomes \cite{bako_unveiling_2024}, the role of visualization example galleries \cite{yang_considering_2024}, and how practitioners view and mitigate design fixation \cite{parsons_fixation_2021}. A better understanding of practitioners' practices relating to inspiration can lead to developing better resources that provide inspiration, and may have implications for how future visualization practitioners should be trained. As the roles for professional visualization designers have been growing in recent years \cite{parsons_understanding_2022,setlur_functional_2022}, and visualization researchers have been increasingly interested in practice-focused research (e.g., \cite{parsons_understanding_2022,alspaugh_futzing_2019,mckenna_design_2014,joyner_visualization_2022,zhang_visualization_2023, yang_considering_2024}), it is an opportune time to investigate phenomena relating to professional design practice.

In this work, we investigated the perspectives of visualization practitioners on design inspiration, aiming to address the following research question: \textit{How do visualization design practitioners use and perceive inspiration in their professional practice?} We contribute an initial understanding of these concepts in a visualization design context, providing opportunities for future research and discourse between researchers and practitioners, where we compare findings with work in other design disciplines. Lastly, we discuss implications for future research and design of visualization tools.

\section{Related Work}
\label{background}
\subsection{Inspiration}
One of the first figurative definitions of inspiration can be found in Oxford English Dictionary as: ``A breathing in or infusion of some idea, purpose, etc. into the mind; the suggestion, awakening, or creation of some feeling or impulse, especially of an exalted kind'' \cite[p. 1036]{simpson_oxford_1989}. Researchers in various domains have investigated the meaning and role of inspiration for several years. From a psychological perspective, Thrash and Elliot \cite{thrash_inspiration_2003} suggested a domain-agnostic conceptualization of inspiration, noting that it has three characteristics: (1) \textit{motivation} is implied within inspiration, (2) inspiration is something that is \textit{evoked} rather than made into existence by chance or act of will, and (3) inspiration involves \textit{transcendence} in the sense that it goes beyond ordinary human actions, concerns, and cognitive processes. In that sense, according to Thrash and Elliot, when a stimulus object \textit{evokes} inspiration in a person, they gain some sort of awareness that goes beyond or \textit{transcends} their regular perspective, and lastly \textit{motivates} them to realize this new awareness into something new. In subsequent work, Thrash and Elliot \cite{thrash_inspiration_2004} argued that the inspiration process has two distinct components, to which they referred as being inspired \textit{by} and being inspired \textit{to}. The more passive nature of inspired \textit{by} consists of a person appreciating the value of some sort of stimulus object (e.g., being inspired by a beautiful flower and appreciating its beauty). On the other hand, the more active nature of inspired \textit{to} consists of a person acting upon this appreciative feeling (e.g., having the desire to design a visualization based on a flower that they saw, as reported in \cite{parsons_fixation_2021}). According to their earlier conceptualization, inspired \textit{by} falls upon the evocation and transcendence characteristics of inspiration, where inspired \textit{to} falls upon the motivation characteristic of inspiration. 

From a design perspective, Gon\c{c}alves and colleagues \cite{goncalves_what_2014} make a similar observation, suggesting that inspiration processes can be more active (e.g., purposefully searching for information) or more passive (e.g., stumbling upon useful information without purposefully searching for it). For example, seeking out temporal data visualizations exemplifies an active inspiration process. In a different scenario, stumbling across a temporal visualization while browsing social media exemplifies a passive inspiration process. In their later work, Gon\c{c}alves and colleagues \cite{goncalves_inspiration_2016} made further distinctions, suggesting that there are 4 types of inspiration processes: active search with purpose, active search without purpose, passive search, and passive attention. Active search with purpose involves a purposeful search for inspiration with a specific goal in mind. For example, looking for specific visualization examples (e.g., treemap) can be an example of an active search with purpose. Active search without purpose refers to a search with similar nature to active search with purpose, but without a specific goal in mind. For example, consulting or going through a common source of visualization inspiration (e,g., visualization blog), without having a specific project task at hand, can be considered as an example of an active search without purpose. Wilson \cite{wilson_information_1997} suggests that this type of search is used to keep one's knowledge refreshed or expanded. Passive search involves an unintentional and random encounter with relevant information that is later integrated into the design problem at hand. For example, encountering an interesting visualization example while browsing social media could be considered as an example of a passive search. Lastly, passive attention involves similar scenario to passive search, but without integration into the design problem at hand. Thus, passive attention is an inspiration that is randomly encountered, but is not immediately used (but could be used in future). In our interviews, we divided these 4 types into two types initially proposed by Gon\c{c}alves and colleagues (active and passive). 

In design, inspiration has been referred to as a ``process that can integrate the use of any entity in any form that elicits the formation of creative solutions for existing problems'' \cite[p. 29]{goncalves_what_2014}. Gon\c{c}alves and colleagues \cite{goncalves_inspiration_2016} conceptualized the inspiration process in design as cyclical and iterative. Typically, the process is initiated by the designer, where afterwards designers use inspirational materials as a starting point. These materials then get adapted to the design process or discarded and this process continues in cycles until the design problem is re-framed or solved. Even though inspiration has been conceptualized and received attention from researchers in various domains, how it is used in professional design practice remains largely unexplored \cite{scolere_digital_2021}.

\subsection{Inspiration and Creativity}
Inspiration has often been discussed in relation to creativity in the design literature. Even though there is no widely agreed upon definition of creativity \cite{silvia_creativity_2010}, most researchers agree that for solutions to be considered "creative", they have to be novel \textit{and} useful \cite{feist_meta-analysis_1998, sarkar_assessing_2011}. Inspiration is essential for creative performance in practically any design profession and researchers have long noted its significance in creativity (e.g., \cite{eckert_sources_2000, goncalves_what_2014, herring_getting_2009, koronis_crafting_2021}), sometimes even using these two words as interchangeable (e.g., \cite{chamorro-premuzic_creativity_2006}). While we do not propose new definitions of inspiration and creativity, we consider these concepts to be distinct and do not use them interchangeably (as some researchers have done in the past). Besides its creativity promoting nature, designers value inspiration for a variety of other reasons. For instance, inspiration alleviates the design process \cite{mete_creative_2006}, triggers idea generation \cite{koronis_crafting_2021, petre_complexity_2006}, reduces resources spent such as time and effort \cite{cai_extended_2010}, expands designer knowledge \cite{goncalves_inspiration_2016}, and helps in communicating designer ideas to other people \cite{eckert_sources_2000, petre_complexity_2006}. 

\subsection{Sources of Inspiration}
In design literature, it is widely accepted that designs are not produced in a vacuum \cite{eckert_adaptation_2003}. Sources of inspiration play an important role in any profession requiring creativity \cite{eckert_sources_2003, yang_design_2005}---which at least some types of data visualization design certainly do \cite{parsons_fixation_2021,bako_understanding_2022,mendez_bottom-up_2017, li_data_2018, dignazio_creative_2017}. Eckert and Stacey refer to sources of inspiration as ``conscious uses of previous designs and other objects and images in a design process'' \cite[p. 524]{eckert_sources_2000}. Gon\c{c}alves and colleagues define a source of inspiration as "any stimulus retrieved from one's memory or from outside world, during (or beyond) a design process, that directly or indirectly influences the thinking process leading up to the framing of the problem or generation of a solution" \cite[p. 3]{goncalves_inspiration_2016}. In that sense, source of inspiration can be anything from a specific visualization example, to a flower or object in nature, to a colleague, friend or other person. 

While many sources of inspiration are encountered externally in the world, designers may also draw inspiration from within themselves. For example, design precedents, defined as ``a designer's store of experiences'' \cite[p. 1]{boling_nature_2021} can act a source of inspiration for a designer in the form of precedent knowledge or experience. The source within the designer can also inspire other designers, where a designer becomes a sort of a source themselves. Scolere \cite{scolere_digital_2021} defined a term ``the digital inspirational economy'' to refer to phenomena where work of one designer becomes an inspiration for another designer. 

Sources of inspiration in design have been studied in variety of disciplines, including textile design \cite{eckert_sources_2000}, fashion design \cite{mete_creative_2006}, industrial design \cite{santulli_introducing_2011}, and interaction design \cite{halskov_kinds_2010}. However, sources of inspiration---and the phenomenon of inspiration in general---have not received similar attention in the visualization community.

\subsection{The Dual Nature of Inspiration}
\label{sec:dual}
Despite the positive connotations of inspiration, many researchers have noted its potential dual nature (e.g., \cite{cai_extended_2010}). While inspiration has been shown to promote creativity and idea generation (e.g., \cite{santulli_introducing_2011, goldschmidt_inspiring_2011}), in some cases inspirational sources can be detrimental to creativity and result in less original solutions. The phenomenon of design fixation, defined as ``a blind, and sometimes counterproductive, adherence to a limited set of ideas in the design process'' \cite[p. 4]{jansson_design_1991}, has been extensively studied in the design literature and replicated across several design disciplines. Originally demonstrated empirically in a classic study by Jansson and Smith \cite{jansson_design_1991}, design fixation often occurs when designers are presented with design examples, where their eventual solutions resemble the features of the examples they saw beforehand. While initial studies showed design students as being vulnerable to instances of design fixation, future studies further demonstrated that design fixation is prevalent among professional and expert designers as well \cite{viswanathan_study_2012, crilly_fixation_2015, condoor_design_2007, kim_design_2014, linsey_study_2010, viswanathan_design_2013}. However, even though there are negative consequences of design fixation, Cross \cite{Cross2001} suggests that it is not necessarily an inherently negative phenomenon in design and that expert designers often engage in fixation-like behavior. Furthermore, all design thinking is influenced by the previous experiences and designs encountered by a designer, and there is no meaningful way to avoid inspirational sources or prevent fixation. Thus it is important to recognize the dual nature of inspirational sources, and to study their nature and use to better prepare future designers and support practitioners in their work.

\subsection{Inspiration in Visualization Design}
Although inspiration per se has not received much direct attention in the visualization literature, researchers have written about inspiration-related practices, tools, and techniques. For instance, Willett and colleagues \cite{willett_perception_2022} have argued that it may be useful to think of the perceptual and cognitive benefits of visualization in a manner akin to the superpowers of heroes. Taking a cue from this notion, they explore how the concept of ``superpower'' can be harnessed to inspire the design of effective visualization systems. Specifically, they describe how fictional superpowers from comic books and related media can act as sources of inspiration for generating new visualization systems. They discuss a variety of visual superpowers that they have collected from various media.

Owen and Roberts \cite{owen_inspire_2023} acknowledge the importance of inspiration for visualization designers and propose a tool called VisDice, aimed at stimulating creative and inspiring ideas. They found that random prompts offered by VisDice encourage creative thinking and help in overcoming mental blockages and fixation. Brehmer and colleagues \cite{brehmer_generative_2022} proposed a technique called Diatoms for generating design inspiration for glyphs. Specifically, the technique samples various elements from a palette of visual objects and generates a variety of options for glyph designs. Besides the proposed technique, the authors also discuss inspiration in visualization design in general, and talk about various sources that designers use including the Data Visualization Society's Slack workspace, social media platforms like Reddit and Twitter, and D3.js or other similar repositories. Judelman \cite{judelman_aesthetics_2004} discussed how visualization research and design is predominantly rooted in computer science and engineering, with visualizations being created largely from people with technically-oriented backgrounds. At the same time, there are multiple resources within the more artistic and design communities that also engage in information visualization, but are being underutilized in primary visualization research. Judelman then proposes and describes various disciplinary sources, such as algorithmic art, architecture, and nature to name a few, where visualization research can draw inspiration from to find new and innovative graphic and interactive techniques for visualization design. He and Adar \cite{he_vizitcards_2017} used inspiration in an educational setting via a design workshop that was conducted in a graduate level course. Specifically, He and Adar used various (including inspiration cards) physical cards, called VizItCards, which were used to facilitate the design process during the workshop with relative success.

Most visualization research broadly related to inspiration has focused on the development of frameworks or tools and techniques that may enhance creativity and inspire. Comparatively less work has taken a practice-focused approach---e.g., investigating the ways in which practitioners seek out, and make use of inspirational sources in their design processes, and the opinions they have towards its relevance in their practice. In one notable contribution, Bako and colleagues \cite{bako_understanding_2022} investigated what types of visualization examples visualization designers use and how they find them. They interviewed 15 students and 15 professional visualization designers and found that visualization designers use multiple diverse sources to find visualization examples (e.g., social media, published media) and employ a variety of techniques and activities (e.g., merging, modifying) to integrate examples into their workflow. In another study, Parsons \cite{parsons_understanding_2022} interviewed 20 visualization practitioners to investigate their design process, specifically in terms of their decision making, methods, and processes. Findings described how designers use precedent knowledge as a source of inspiration, and reported some of other sources that designers tend to use (e.g., nature). In their later work, Parsons and colleagues \cite{parsons_fixation_2021} investigated perceptions of visualization designers on design fixation, specifically factors that designers think encourage or discourage fixation. The findings described some of the inspiration strategies that designers use to avoid fixation (e.g., looking at sources that were not closely related to problem at hand) and also reported on some of the sources that visualization designers use (e.g., existing visualizations, art, nature). 

While there are several contributions in the visualization literature related to inspiration, broadly construed, we still know very little about how visualization practitioners perceive, seek out, and use inspiration sources, and how they relate to other aspects of their practice. Inspiration is essential for any design profession; designers regularly seek out sources of inspiration, and also draw on precedent designs and other forms of design knowledge that serve as conceptual and creative springboards for design activity \cite{leifer_early_2014, eckert_sources_2000, koronis_crafting_2021}. While some types of visualization design (e.g., business reporting) may be more routine and less creative (e.g., relying on pre-determined templates and visualization types), it is not unusual for visualization design to be more open-ended, requiring creative solutions that must inevitably draw inspiration in some form.

\section{Method}
To understand the nature of inspiration in visualization practice, we recruited participants who self-identified as professional data visualization practitioners. Recruiting was done via social media, the Data Visualization Society's Slack workspace, and our personal networks. To mitigate sampling bias, we advertised widely and directly contacted more than 100 practitioners. 

Interviews followed a semi-structured protocol and were all conducted remotely via Zoom. We started the interviews by asking participants to describe some projects they had worked on, and asked follow up or clarifying questions related to their inspirational practices and beliefs based on the answers given. After that, we asked specific questions regarding their thoughts on the concept of inspiration itself, such as what they think inspiration is, what their views on inspiration are, whether they thought of inspiration as important to their design process, and what factors affect (if any) their inspiration practices and process. We also asked questions related to sources of inspiration, such as what kind of sources they use, whether it is easy for them to find inspiration and inspiration sources, and what the criteria are for "good" sources. We asked whether designers consider their inspiration process to be active, passive, or both. Lastly, we asked their thoughts on concerns with accidentally copying work from inspiration sources and effects of inspiration on their sense of ownership. All interviews were conducted by the first author.

\subsection{Participants}
In total, we had 14 participants who self identified as data visualization practitioners. Most participants had between 2 and 5 years of professional experience, with only few participants (P1, P2, and P8) having less than 2 years of professional experience and only two participants (P4 and P6) who had more than 5 years of professional experience. In terms of work roles and types of visualization work, we had a fairly diverse set of participants, ranging from newspaper and journalism contexts (e.g., P1, P2) to internal dashboards (e.g., P3) and print media (e.g., P12). The industry in which the participants work is presented in Table \ref{tab:demographics}. We also present information about the projects that participants described during the interview to give context to their responses. This information is deliberately abstracted to protect participants' identity. 

\subsection{Positionality}
As researchers, we bring diverse disciplinary backgrounds and professional experiences to this study. Our backgrounds span computing, design, and cognitive psychology, with over 15 years of combined experience studying design practices and processes. Two of the researchers have direct experience in developing visualization systems, including one who led the development of a visual analytics tool deployed in a commercial setting. These experiences provide us with practical insights into the challenges and nuances of visualization design, enabling us to approach this study with a deep appreciation for the complexity of the practice. At the time of conducting this study, the authors had no personal or professional conflicts with visualization practitioners, companies, tools, or their developers. 

Our perspectives are rooted in the belief that inspiration in design is not merely a cognitive phenomenon but a dynamic interplay of individual experiences, contextual factors, and creative practice. This belief shapes the way we interpret the participants’ insights and situates our analysis within a broader understanding of design as an inherently situated human-centered activity. We are committed to practice-focused inquiry, foregrounding the perspectives and lived experiences of practitioners over controlled experimental approaches. This commitment reflects our view that the realities of design work are best understood by engaging directly with those who practice it, capturing the richness and nuance of their processes.

We recognize that our disciplinary lenses, professional experiences, and practical backgrounds inevitably influence the research process, from the formulation of research questions to the interpretation of findings. By acknowledging these influences, we aim to foreground the voices and lived experiences of the designers in this study while also reflecting critically on how our positionality shapes the narrative we construct. 

\subsection{Data Analysis}
Interviews ranged from 42 to 74 minutes, with an average length of 56 minutes. The interviews were initially transcribed using automated transcription tools, followed by manual error correction. We performed a hybrid thematic analysis, involving both a top-down, deductive, theory-driven process and a bottom-up, inductive, data-driven process \cite{fereday_demonstrating_2006}. This type of hybrid approach allows for exploring layered and complex phenomena that can benefit from both the deductive application of concepts from established theory, and the inductive generation of new and interesting topics that may not already be established in existing frameworks. The topic of inspiration, while grounded in some extant literature, has received little attention in the context of visualization. This dual approach allowed us to leverage the existing body of knowledge while remaining open to identifying novel and unexpected insights relevant to professional visualization practice.

We followed the three-phase model outlined by Swain \cite{swain_hybrid_2018}, which builds on the hybrid approach described by Fereday and Muir-Cochrane \cite{fereday_demonstrating_2006}. This three-phase model provides a flexible yet systematic framework for hybrid thematic analysis. All three authors participated in the analysis, with the third author serving a supervisory role. First, we familiarized ourselves with the data by reading the interview transcripts and reflecting on the literature outlined in Section \ref{background}. This immersion helped us identify key concepts and themes from the literature, which were used to establish our initial set of a priori codes. These included: sources of inspiration; active and passive forms of inspiration; and positive and negative perspectives on inspiration. 

Second, we applied the a priori codes to the transcripts while simultaneously generating a posteriori codes through inductive analysis. This process was iterative and collaborative. The first two authors split the transcripts and conducted an initial round of independent coding, focusing on both deductive and inductive themes. The team then met to discuss the application of the a priori codes, ensuring consistency and alignment across coders. Given the conceptual clarity of the a priori codes, their application was straightforward, with clear agreement observed among the coders. As a result, we did not see a need for inter-coder reliability scores or other similar measures, which can be an overkill in such situations \cite{mcdonald_reliability_2019}. The inductive coding process, while data-driven, was not entirely atheoretical, as it was inevitably influenced by our prior knowledge and experience with design theory and visualization practice. These inductive codes were iteratively refined through team discussions, ensuring their relevance and alignment with the research focus. Codes and notes were managed in Dovetail\footnote{https://dovetail.com/}, allowing for systematic tracking and collaborative review.

Third, we reviewed the codes to identify overarching themes we thought represented the  central insights. A priori and a posteriori codes were examined for conceptual overlap and combined into higher-level themes where appropriate. Our initial coding  led to the creation of 88 codes organized into 8 themes. We also modified the initial a prior codes to be higher-level themes, with inductively derived codes coming underneath. We share the same position as Swain \cite{swain_hybrid_2018}, that the difference between a code and a theme is not of much importance. We treat them as conveying different levels of abstraction, but not as indicating any fundamental difference in quality or type of knowledge. Through discussion and iterative review, we collapsed the codes down to 56, organized under 4 themes. This organization formed the basis of what is presented in this paper. These themes reflect both the theoretical foundations of the study and emergent insights from the data, offering a nuanced understanding of how inspiration is perceived and utilized in professional visualization practice.

\begin{table*}
  \caption{Demographics of our study participants and project context.}
  \label{tab:demographics}
  \centering
  \begin{tabular}{cccc}
    \toprule
    Participant & Industry & Years of Experience & Project Context \\
    \midrule
    P1 & Journalism & 0 - 1 & Analysis of people's representation in currency   \\
    P2 & Journalism & 0 - 1 & Story on climate change\\
    P3 & IT & 2 - 5 & Dashboard (IT infrastructure) \\
    P4 & Freelance & 11+ & Sports car racing data \\
    P5 & Consulting & 2 - 5 & Mockup tool for data analysts and BI developers\\
    P6 & Freelance & 6 - 10 & Dashboard (Risk Management)\\
    P7 & Finance & 2 - 5 & ``Scrollytelling'' data story on China\\
    P8 & Finance & 0 - 1 & Project on cybersecurity\\
    P9 & Freelance & 2 - 5 & Impact of tech on reading\\
    P10 & Freelance & 2 - 5 & Geographic and emotional visit maps\\
    P11 & Education & 2 - 5 & Algorithmic fairness\\
    P12 & Freelance & 2 - 5 & Guide for safe spaces\\
    P13 & Education/Journalism & 2 - 5 & Global and domestic aviation network\\
    P14 & Freelance & 2 - 5 & Social Media Mental Health Support\\
  \bottomrule
\end{tabular}
\end{table*}

\section{Findings}
We organize the findings by the themes defined during the coding process. The themes are as follows: Where Designers Find Inspiration, Importance of Inspiration in Design, Inspiration Practices, and Balancing Imitation and Sense of Ownership.

\subsection{Where Designers Find Inspiration}
Participants described a variety of sources of inspiration. Most unsurprisingly, most participants reported using sources like social media (e.g., P1 - \textit{``it's usually Instagram or Twitter. I'll look for hashtag data visualization and just look for stuff.''}; P5 - \textit{``... lot of these design sites like Dribbble, Behance ... even Pinterest at times ... provide a lot of inspiration to me.''}), built-in recommendations from visualization tools (e.g., P14 - \textit{``...but a lot of them, you know, they have these built in chart types and things like Power BI and Tableau.''}), and data visualization blogs and catalogs. (e.g., P9 - \textit{``some of the blogs and the, at least the digital version of the Data Visualization Society Nightingale magazine. Yeah, that's a big source of inspiration as well.''}; P11 - \textit{``Yeah, Information is Beautiful ... absolutely incredible source of visualization.''}). Arguably less obvious (but more prevalent in other design disciplines) sources of inspiration included art exhibitions (e.g., P1 -  \textit{``... a new approach that I have been getting from attending art exhibitions that are visualizing data.''}), museums (e.g., P8 - \textit{``...but oh, this one's at a museum. I love this. This was at the science museum in Boston.''}), nature (e.g., P12 - \textit{``...as I was taking a walk in my neighborhood, I saw an old tree stump... ...And so I just came up with the idea there, what about if we represented surveys and the scope of surveys through a tree trunk.''}), and abstract concepts (e.g., P7 - \textit{``And because I have read somewhere that in Asia, time was seen as something cyclical.''}). More surprisingly, several participants considered individual people (sometimes even themselves) as sources of inspiration. Interestingly, participants did not reference specific pieces of work; rather, they simply referenced the people themselves as sources of inspiration. For example, P1 stated \textit{``you've probably heard of this girl Mona Chalabi. I find her inspirational---just like her, as a person.''} P8 described using their own work as inspiration: \textit{``Even my work, like this is my work... so this being able to archive these works is always helpful to have fun references [for future inspiration].''} P10 remembered instances where other people were inspired by their work: \textit{``I'm not super famous or whatever, but it happens once or twice that I remember that some students or other people, like practicing in data visualization, told me: I love your work, so I looked at them [for inspiration].''}

\subsection{Importance of Inspiration in Design}
There was pervasive sentiment among participants that inspiration plays a crucial role in the creative process and is generally a positive phenomenon. For example, P14 described inspiration as a positive feeling, even suggesting that if they were to have this feeling all the time, they would feel great: \textit{``If I was feeling inspired all the time, I would feel great.''}. Some of the other participants had similar perceptions about inspiration, with P12 considering inspiration as a motivation booster (e.g., \textit{``When I'm inspired, I feel more motivated to do my work.''}), P2, P4, and P5 noting the practical utility of inspiration (e.g., P5 - \textit{``[Inspiration] It adds a lot. It gives you a starting points for you to work from ... if I start working on a project and starting from scratch, if I don't have an inspiration, or I don't have say a point of reference that I want to go back to, it becomes difficult for me to actually put my thoughts down in terms of design, but when I see something that I like or ... see some something that inspires me, it just gives that sort of a kickstart to me that okay, this is the direction I can go on.''}), and P2 emphasizing a sense of community through the acts of being inspired by other designers and inspiring other designers themselves. On the other hand, one of the participants (P9) expressed a potential negative aspect of inspiration, where they described that sometimes looking at other people's work makes them feel insecure \textit{``when I am going through a phase of insecurity ...  in those moments, it would make me go, you know, I really don't know how to do much and feel bad that I haven't made such a contribution to the world.''} Even some of the more experienced participants (e.g., P4) voiced similar concerns: \textit{``... possibly there is sometimes too much inspiring work [that] can lead ... to feeling inadequate.''}

\subsection{Inspiration Practices}
\subsubsection{Types of Inspiration Processes} \hfill\\
\textbf{Active or Passive} Participants were fairly spread out in terms of whether their inspiration process was active, passive, or a mixture. Some participants described specific moments where their process switched between active and passive. P10 stated: \textit{``If I have to look for references for a specific project, I would go for the active, inspiration stuff, but also through the time, always looking for references. It's kind of mindset.''} P13 discussed the influence of their co-workers on their process: \textit{``I used to be quite passive and now I'm more active... ...and I think it's the influence of the designers on our team.''} P8 talked about their process as a feedback loop (\textit{``...it's a feedback loop ... my active inspiration is going to affect the quality of my passive inspiration.''}), where if they were looking for inspiration purposefully, it would later keep them alert for passive instances of similar inspiration, and vice versa when they randomly encountered a relevant inspiration piece, they would later explore that area more purposefully by looking for related inspiration material.

\textbf{Active Inspiration} Several participants described instances where they actively looked for inspiration. P13 stated: \textit{``I felt that my GIS, my mapping skills were extremely lacking. So I did a deep dive into some of the famous examples of mapping and cartography.''} P2 mentioned: \textit{``Because, for example, looking for data visualizations, for the climate displacement work project, I was actively looking for, a there was this time where every day I went to Climate Lab of the Washington Post.''}

\textbf{Passive Inspiration} Some participants described instances of passive inspiration. P2 stated: \textit{``And other times, it's passive, for example, the Data Vis Dispatch is something that is every week, and I look at the all the data visualizations, not looking for something concrete or something special.''} P5 mentioned: \textit{``...it usually happens when I'm  scrolling through social media ... I'm looking at a Youtube short or an Instagram reel, it will be completely unrelated to data visualization or anything that I do.''} P4 talked about getting an idea for a project while watching a show: \textit{``And in terms of inspiration ... it was actually the Netflix series ... which got me interested in the topic.''}

\subsubsection{Inspiration Activities} \hfill\\
Participants engaged in various types of activities with the inspiration that they encountered. \\
\textbf{Curation} Many participants mentioned collecting inspirational materials for future use. P8 stated: \textit{``I mean, it is part of my routine ... I am a collector by nature, I like to collect inspiring pieces ... and even my work that's inspiring.''} P10 discussed their strategy for collecting inspirational material: \textit{``... when I see something interesting, I just save it or do a screenshot.''} P4 mentioned: \textit{``I like gathering inspirational imagery or other interactives that I liked.''} \\
\textbf{Transformation} Some participants used inspiration materials and transformed them into something new for their own project. For example, P4 described using a plot: \textit{``... having seen the plot from the data set, on Kaggle, it was basically turned 90 degrees, it went from left to right. And I thought it, if I twisted it 90 degrees, it kind of looked like lightning.''} \\
\textbf{Replication} Some participants discussed how they replicated some elements from inspirational materials, often for educational purposes. P2 discussed their strategy: \textit{``I got to Twitter or go to New York Times, Climate Lab and start seeing how other people use this kind of data... ...and while doing the data analysis and data exploration, I try to replicate it in a very basic way, not adding interactivity and add color.''} P1 also described how they sometimes try to emulate the style of the work that they liked: \textit{``... sometimes ... just be the type of dataviz that I really like on someone's article, and then I am like ... maybe I should use that one ... [and] how can I incorporate that style with my work or data that I am using''} \\
\textbf{Sharing} Some participants also discussed how they shared inspirational materials with their co-workers. P2 described how they shared a source with their co-worker: \textit{``And actually, I, every Tuesday, I sent him the Data Vis Dispatch from Datawrapper, and we talked about the data visualization that they showed.''} P10 mentioned how they created and curated a collaborative inspiration folder in Google Drive and often used inspirational materials stored there in discussions: \textit{``And we saved all the ideas and references into a folder [in] Google drive ... We just collected ideas, we had folders by date. [Then] hey, this is my most recent research. What do you think about it? And I think it's too playful. Okay, let's do another one.''}

\subsection{Balancing Imitation and Sense of Ownership}
\subsubsection{Imitation versus Inspiration} \hfill\\
Many participants expressed concerns with accidentally copying from or imitating work that inspires them. P14 mentioned how using one source of inspiration can lead to such concerns: \textit{``And if it's like, I have only had one inspiration source and I am doing one thing, then that's kind of a recipe for disaster.''} P12 also described how they consciously try to avoid copying when looking at other people's work: \textit{``And so I think I am just trying to also avoid copying as well.''}. Majority of participants, however, had no issues or concerns with copying at all, usually accepting that copying in some form is part of the job. P9 commented how copying and remixing good examples helps them learn: \textit{``Because I really think one of the best ways of learning is copying at the people's work and then, kind of remixing and, you know, later coming up with your own ideas.''} P10 argued that copying in some form is part of a designer's job: \textit{``Designers do copy... ...taking inspiration, it means copying right. It is like artists create stuff, designers copy.''} P3 also suggested that work conditions encourage imitation in some cases, where clients might strongly encourage or even demand that designs look a certain way: \textit{``I mean, as a designer, you always want to do something that's brand new and creative... ...but a lot of times your clients and your users, they want something that's more familiar.''}

\subsubsection{Sense of Ownership} \hfill\\
Participants discussed and expressed their thoughts on the nature of the work ownership in relation to inspiration, where they felt that even an inspired work is still uniquely their own. P12 discussed how they think the final work is still original if they were inspired only by a certain part of the inspiration piece: \textit{``I think it becomes my own if I am inspired by an aspect and not an overall visual look.''} P11 also described how any designer will inevitably change the inspiration piece anyway and look at it from a different perspective than the original author, which adds to the sense of ownership: \textit{``And so you are always gonna have to be adding things or changing things or maybe you are thinking through an accessibility component that, that previous person was not thinking through.''} P10 expressed similar thoughts, where they thought that due to personal preferences and styles, the work becomes their own by default: \textit{``I also think it is really difficult to do the same stuff, if you change the shapes or dataset itself. Also personal style every time just, comes up, it is like natural.''}

\section{Discussion}
Building on the findings above, we discuss the diverse sources of inspiration designers draw upon, such as academic resources, AI tools, peers, and visualization software, and their accessibility and relevance. We also discuss designers’ practices, such as curation, transformation, and replication, and how these align with or diverge from findings in prior research. Finally, we suggest how existing visualization frameworks, with their limited focus on inspiration and design cognition in general, could be strengthened by incorporating insights about inspiration. By linking our findings to prior work, we suggest opportunities to enhance both theoretical understanding and practical support for inspiration in visualization practice.

\subsection{Inspiration and Other Design Knowledge Beyond Visualization Examples}
Visualization designers regularly engage with problems that cannot be sufficiently guided by well-defined design models and frameworks \cite{parsons_preparing_2023}. These kinds of design problems require the integration of several additional types of knowledge, which may involve precedent artifacts and various forms of inspiration. Our work demonstrates not only that inspiration is an important element of visualization designers' work process, but also that visualization designers highly value inspiration and often actively seek it. Visualization researchers have recently recognized the value of forms of design knowledge beyond the classically articulated ones---such as frameworks and models, encoding and interaction techniques, taxonomies and typologies, and design principles and guidelines---turning their attention to the role of visualization examples in design practice (e.g., \cite{bako_unveiling_2024, bako_understanding_2022, yang_considering_2024}, c.f. \cite{parsons_what_2020, parsons_considering_2022}). While visualization examples can serve as important sources of inspiration, they are only one type of inspirational source. Participants in our study described using a variety of sources that go beyond visualization examples, including sources that have no direct relation to visualization or data, including art exhibitions, nature, abstract concepts, movies, and so on. Furthermore, although visualization designers sometimes seek out visualization examples as sources of inspiration, other times they actively avoid visualization examples and seek out other sources of inspiration instead (see e.g., \cite{parsons_fixation_2021}). Our findings offer an expansion on the view of what useful sources of inspiration might be for visualization practitioners. While the investigation of visualization example galleries is important and relevant work \cite{yang_considering_2024}, future research should also investigate the curation of other types of galleries. For instance, there may be types of art, philosophies of art, or even periods and movements within art history that could inspire visualization design in different ways. These various sources could be studied and curated into galleries. Similar points can be made about the other sources of inspiration mentioned by our participants, including objects and processes in nature, movie styles, and abstract concepts. Despite knowing that such phenomena are valuable for visualization design, we know little about how they can be structured to inspire practitioners in helpful ways.

\subsection{Sources of Inspiration}
\subsubsection{Academic Sources}
Notably, participants did not mention using any resources that have been generated by academics (e.g., the several online catalogs of visualization techniques \cite{schulz_treevisnet_2011, kucher_text_2015, von_landesberger_visual_2011}). This finding, while interesting, is not entirely surprising, as the disconnect between academic research and design practice has been noted to some degree in the visualization community (e.g., \cite{parsons_what_2020}) and especially within the larger human-computer interaction (HCI) community (e.g., \cite{colusso_translational_2017}). Still, researchers could investigate this issue further to understand why such catalogs are not being consulted. For instance, it would be valuable to investigate whether this is simply an issue of visibility, which could be relatively easy to fix. Or, perhaps there is a more fundamental disconnect between researchers and practitioners that could be surfaced and studied. Furthermore, such efforts could even lead to a deeper understanding of inspiration and how it intersects with notions of relevance, belongingness, and trust.

While improving existing inspiration resources was not the purpose of this work, our findings suggest some directions that future work on inspiration resources could take. Many visualization designers from our study described their inspiration process as more passive (e.g., stumbling upon a potentially relevant information rather than purposefully searching for it), which could suggest that building resources around and for such practice can be beneficial for visualization design. One of the participants (P14) described a browser plugin that they use, which shows them various design related information every time they open a new browser window. Popular visualization tools like Tableau could consider ways to support this type of passive inspiration---perhaps even finding ways to include sources not directly related to visualization examples. Several designers also mentioned collecting inspiration materials for future use and sharing pieces with their colleagues, friends, or employers. Supporting such activities within the tools and resources (e.g., similar to Tableau's vis of the day or various galleries) can also be beneficial for inspiration practices.

\subsubsection{AI Sources}
None of our participants mentioned using any artificial intelligence (AI) generated resources for inspiration. Given the rising popularity of generative AI and the widespread access to tools like ChatGPT, it is somewhat surprising that practitioners did not discuss any AI tools or AI-generated resources. Recent trends in the visualization community suggest benefits of generative AI models for visualization research and design (e.g., \cite{shi_supporting_2023, xiao_let_2023}). While benefits of text-to-text AI models (e.g., ChatGPT) for inspiration purposes are debatable, there may be potential of text-to-image AI models (e.g., Midjourney) used as inspiration. However, this is a tricky issue, as the hype around LLMs appears to be waning as their limitations for serious creative work are made more clear over time. Future studies could examine how and under what conditions AI generated images may serve as sources of inspiration rather than generators of usable visualization outputs.

\subsubsection{People as Sources}
Our participants often referenced other people as sources of inspiration, often even as a separate entity from their work. Scolere \cite{scolere_digital_2021} discussed inspiration within a digital economy of design inspiration platforms (e.g., Behance), where inspiration acted as a currency between different designers, where one designer's work can act as inspiration for another designer. Based on our findings, a similar phenomenon may be happening in the visualization practitioner community, however some of the visualization designers we interviewed also described prominent figures within the community as sources of inspiration and not necessarily their work, but themselves as a separate form of inspiration source. Some participants even referred to themselves as sources of inspiration. However, compared to Scolere's notion of inspiration currency, visualization designers were sometimes inspired by their own previous work, rather than other designers, through precedent knowledge or curated inspiration collections for future use. Many researchers have reported on similar phenomena in architecture with research on architects' precedent knowledge and experiences \cite{gero_design_1990, oxman_prior_1990, schon_designing_1988}, as well as research on how architects' own work in the form of sketches stimulated their idea generation \cite{goel_sketches_1995, goldschmidt_dialectics_1991, goldschmidt_serial_1992, schon_reflective_1983, schon_kinds_1992, purcell_drawings_1998}. The concept of self-sourcing inspiration is certainly interesting and warrants further investigation, as curating inspiration materials (including one's own work as a potential future inspiration source) seems to be a fairly common practice. It would be valuable to study how  using one's own (versus others') work as inspiration influences creativity and design fixation. Research in other design domains shows that experts rely on patterns from their prior experience to make good judgments (or "gambits") that help them connect problems to solutions \cite{lawson_schemata_2004}. Perhaps gambits are being recognized and cataloged when visualization designers refer to themselves as sources of inspiration, although this is only speculation on our part at this stage of the research.

\subsubsection{Visualization Tools as Sources}
Visualization tools ranging from Excel to Tableau and D3.js have chart or encoding presets either built within the tool itself or through curated community libraries and galleries built around the tool. Designers described these presets acting as a source of inspiration---however, often in a more negative sense rather than a positive one (as discussed in Section \ref{sec:dual}). Bako and colleagues \cite{bako_understanding_2022} reported that designers tend to not use the built-in chart recommenders much in their practice. Parsons and colleagues \cite{parsons_fixation_2021} found that practitioners recognized chart recommendations as constraining their creativity and potentially leading to fixation. M\'endez and colleagues \cite{mendez_bottom-up_2017, mendez_considering_2018} investigated how such presets (and other forms of automation) can be optimized to increase the efficiency of the design process while maintaining user agency and creativity. There are tradeoffs in providing such recommendations, as constructing visualizations from the bottom up requires more thoughtful engagement with the problem and more creativity in the output \cite{mendez_bottom-up_2017}. This phenomena of having software-recommended solutions is perhaps not as common in other design disciplines---as there may be more freedom deciding on materials and forms in the solution space---while visualization recommendations can be made automatically based on data types, tasks, and known encodings (e.g., as in \cite{moritz_formalizing_2019}). This may be a topic that is interesting not only to visualization researchers and practitioners, but also to the design community more broadly as design presets as inspiration are not as common outside of the visualization design. Researchers can investigate if giving different types of design presets can result in more novel and original solutions.

\subsection{Practices Relating to Inspiration}
\subsubsection{Active and Passive Inspiration}
Many of the inspirational practices that visualization designers engage in have been in line with previous work in design and visualization literature. However, while some of the researchers noted the more active inspiration approach of designers (e.g., \cite{eckert_sources_2003}), our findings suggest a more evenly spread out approaches of visualization designers with many of them engaging in a more passive approach in their inspiration practices. Given the frequency of visualization designers engaging in passive inspiration activities, researchers could investigate ways to support passive inspiration by perhaps creating some sort of daily/weekly inspiration mail lists, catalogs, or websites that practitioners could subscribe to. 

In the visualization community, Sprague and Tory \cite{sprague_exploring_2012} explored casual encounters with visualizations and among other findings proposed several design implications for designing visualizations better suited for casual encounter and use. While casual encounters and passive inspiration/attention are similar in spirit, passive inspiration/attention (and inspiration in general) implies some sort of action (now or in future) based on the stimulus object (see motivation in Thrash and Elliot's \cite{thrash_inspiration_2003} conceptualization of inspiration described in Section 2.1), while casual encounter does not necessarily imply any action at all. Furthermore, compared to the study by Sprague and Tory, our participants are visualization practitioners, thus there are significant differences in visualization literacy and expertise, suggesting that some design implications will not easily translate to more practitioner-focused resources. 

\subsubsection{Inspiration Activities}
Some of the inspiration activities of visualization designers have also been reported in other design domains, such as interaction design (e.g., \cite{halskov_kinds_2010}) and also within the larger HCI as well. Specifically, Halskov reported interaction designers engaging in activities similar to replication and transformation activities that we found. Zhang and Capra \cite{zhang_understanding_2014} also reported that their participants engaged in transformation activities in their creative tasks. Interestingly, Bako and colleagues \cite{bako_understanding_2022} reported that only half of their participants engaged in some form of curation in relation to visualization examples, while curation seemed to be more prevalent in our findings. Furthermore, curation appears prevalent in other design and HCI domains as well (e.g., \cite{keller_collecting_2009, keller_collections_2006, kelley_art_2001, kang_paragon_2018, kerne_strategies_2017}). Collecting and curating inspiration pieces seems to be an important activity in a design process which is also applicable to visualization design based on our findings. Our participants organized their collections by keeping folders of their past work, screenshots of relevant objects, works of other people, and so on. The benefits of such collection and curation activities are not quite clear and researchers could further investigate whether there are any significant differences in performance and in novelty of solutions between seeking inspiration through one's own collection or doing a more impromptu search. Furthermore, as such activities are fairly prevalent among designers, future research could investigate how to better support such activities in relation to new or existing practitioner-oriented resources.

\subsection{Inspiration, Imitation, and Professional Identity}
Our participants shared thoughts on copying other people's work and design fixation. Scolere \cite{scolere_digital_2021} discussed similar issues in their study of graphic designers, where they referred to these issues as tensions between inspiration and imitation. While Scolere found more negative sentiments towards imitation prevalent among graphic designers, visualization designers in our study were often less concerned with imitation, and even often accepted some copying and imitation as part of the profession. This acceptance of copying may be due to influences of software development, where reuse of code has long been an accepted practice, as suggested by Yang et al \cite{yang_considering_2024}. At the same time, visualization designers often expressed desires to be unique and have their own voice or style in their work. This desire aligns with findings in other disciplines relating to professional identity formation, in which developing a sense of personal style and design philosophy is essential \cite{watkins_tensions_2020, nelson_design_2002}. This is particularly relevant for data visualization at this point in time, when the formation of a collective professional identity is emerging \cite{parsons_understanding_2022, meeks_2018_2018}. Future work should investigate this interplay between reuse, personal style, and professional identity. There may even be aspects of this intersection that are distinct to data visualization, as it is a practice that combines software engineering, data science, and aesthetics in ways that may be unique among design disciplines. 

This tension presents an interesting challenge and opportunity for inspiration researchers, especially in the context of design fixation. Classical design fixation studies often assess performance of designers in terms of their solutions being novel and original; however, subjective experiences and perspectives of designers can present new avenues for research in this area. While metrics like novelty, quantity, and diversity of design outputs are valuable in relation to creativity and fixation (e.g., as in \cite{bako_unveiling_2024, jansson_design_1991, purcell_design_1996}), the personal and subjective aspects of professional practice, including professional identity and design philosophy, will undoubtedly impact the actual practices around inspiration in the wild. This highlights the importance of engaging in practice-focused, naturalistic research in addition to the more controlled and experimental work that is commonly done in visualization \cite{parsons_understanding_2022}.

\subsection{Visualization Design Models and Frameworks}

A variety of frameworks and models have been developed to describe the design process and provide researchers and practitioners with advice and guidance (e.g.,the Nested Model \cite{munzner_nested_2009} and its Blocks and Guidelines extension \cite{meyer_nested_2015}). Well-known process models include the nine-stage framework in the Design Study Methodology \cite{sedlmair_design_2012}, the Design Activity Framework \cite{mckenna_design_2014}, and others \cite{goodwin_creative_2013, mccurdy_action_2016, sedig_design_2016}. Our work here, and others in recent years \cite{parsons_design_2023, parsons_understanding_2022, bako_understanding_2022, bako_unveiling_2024, bressa_sketching_2019, xia_dataink_2018, roberts_sketching_2016, parsons_design_2020} suggest there are essential aspects of the design process and design cognition that are not accounted for, such as the role of inspirational sources, visualization examples, and other forms of precedent and intermediate-level knowledge \cite{parsons_what_2020, hook_strong_2012}. While these extant frameworks are generally created by and for researchers, there is still an opportunity to incorporate aspects of  inspiration into them, as it is probable that researchers also employ such types of knowledge while designing. These forms of design knowledge are not peripheral or optional; rather, these recent studies demonstrate how they are core aspects of the decision making and everyday activities of visualization practitioners. While extant visualization process and decision models are still be valuable in several ways (e.g., for guiding visualization design studies \cite{sedlmair_design_2012}, for anticipating common pitfalls \cite{munzner_nested_2009, meyer_nested_2015}, and describing the integration of process and decision activities \cite{mckenna_design_2014}), expanding on this body of literature to account for the critical role that inspiration and affiliated concepts play would be a valuable contribution to the theoretical frameworks pertaining to visualization design broadly construed, and could augment contributions made in the form of visualization `design studies' \cite{sedlmair_design_2012}. Furthermore, while visualization researchers have become increasingly interested in professional practice in recent years, integrating inspiration into descriptions of design activities (e.g., finding and using examples \cite{bako_understanding_2022}, negotiating with clients \cite{lee-robbins_client-designer_2024}, shaping handoffs to developers \cite{walny_data_2020}, and using writing to guide early design stages \cite{stokes_its_2024}), would strengthen our understanding of professional visualization practice.    

\subsection{Inspiration and Beyond}
Inspiration has received considerable attention within the larger HCI and design communities, where researchers investigated issues related to inspiration within a variety of design domains. Similar to our findings, several researchers have confirmed and acknowledged the importance of inspiration in practically any design domain, and have proposed and tested various abstract inspiration tools to promote creativity in the design process. For example, Halskov and Dalsgaard ran a series of inspiration card workshops where designers had to come up with design concepts using a variety of idea-generating abstract physical cards \cite{halskov_emergence_2007, halskov_inspiration_2006}. Lucero and Arrasvuori \cite{lucero_plex_2010} used a similar relatively abstract tool, also in the form of physical cards (PLEX cards) to design for playfulness. While our participants described using a variety of resources as inspiration in their design process, none of them described using abstract, multi-purpose, and dedicated inspiration tools in the form of physical cards. Attempts have been made to investigate the value of inspiration cards for teaching students about visualization design \cite{he_vizitcards_2017}. Future research is needed to investigate the utility of multi-purpose inspiration tools for professional visualization design.

While research in the visualization community has investigated inspiration largely through the lens of examples and prior work, our findings have tried to expand this lens to broaden how inspiration is construed in visualization research. Similar efforts have been made within the larger HCI community, for example where Miller and Bailey investigated how designers search for inspiration in the form of examples, construing examples as ``any item that inspires creative thought'' \cite[p. 1]{miller_searching_2014}. Our work found that visualization designers described a variety of sources they found inspiring beyond visualization examples and prior work. Future treatments of these topics should expand on the notions of inspiration, characterizing the nature of design knowledge more broadly in data visualization practice---a topic that has not received much research attention to date. Researchers investigating other disciplines of design practice have characterized inspiration and examples as just some of many types of design knowledge, along with other forms of precedent knowledge like lived experiences stored as episodic memory \cite{Visser1995, lawson_schemata_2004}, and prototypical examples of design approaches  \cite{boling_nature_2021}. Our understanding of data visualization design would benefit from richer descriptions of design knowledge and how inspiration fits within its landscape.

\subsection{Limitations}
In this study, we conducted semi-structured interviews with 14 practitioners to explore their perceptions and beliefs about inspiration in design. As with any study, there are limitations and caution should be exercised in interpreting the results. For example, the sampling strategy, which relied on convenience sampling, may have been subject to a known sampling bias, as participants who volunteer are often more invested in the topic and may hold strong opinions about certain outcomes \cite{moore_statistics_2006}. Additionally, as the data relied on self reports from our participants, they are subject to biases of retrospective accounts and likely do not fully capture the actual behaviors and attitudes of visualization practitioners. To enhance the comprehensiveness of our research, future studies could complement this approach with alternative methodologies like direct observations, diary studies, or the use of design probes. The relatively smaller sample size of our participants may limit the representation of diverse perspectives within the visualization designer community. Lastly, most of our participants reported low to medium levels of expertise. As prior work on novices and experts shows, it is likely that those with high levels of experience employ different strategies and behaviors. As a result, our findings may be more representative of practitioners with medium or lower levels of experience. However, as the professional identify of visualization practitioners is still very much in flux, we believe this work still provides meaningful insights into the current state of visualization practice.

\section{Summary}
Virtually all design outcomes emerge from some combination of previous designs, phenomena in the world, and the personal experiences and knowledge of the designer. Design rarely proceeds `from scratch', and almost always involves one or more sources of inspiration. While researchers across several design disciplines have studied the nature of inspiration in professional practice, little is known in relation to visualization design practice. Understanding how designers perceive, seek out, and use sources of inspiration has several benefits to researchers, practitioners, and educators, including identifying ways to better support creativity and mitigate fixation, determining which kinds of inspiration produce better design ideas, providing vocabulary for thinking about new designs and describing designs to others, and better understanding how idea generation happens. Our findings from interviewing visualization practitioners offer several paths forward, some in ways that mirror findings in other design disciplines, and others in ways that open new opportunities for contributing to discourses on design inspiration more broadly. We hope our work inspires more practice-focused visualization research and draws more attention to inspiration and related phenomena in professional practice.

\begin{acks}
This work was supported by NSF award \#2146228. We would like to thank our participants for their time and input into this study.
\end{acks}

\bibliographystyle{ACM-Reference-Format}
\bibliography{references}


\end{document}